\documentclass[aps,preprint,superscriptaddress,showpacs,amssymb,amsmath,amsfonts,floatfix]{revtex4}
\usepackage{graphicx}

\begin{document}

%
\title{Absolute Total $np$ and $pp$ Cross Section 
       Determinations}

\newcommand*{\GWU}{Center for Nuclear Studies, Department of Physics, \\
            The George Washington University, Washington, DC 20052, USA}
\newcommand*{\NIST}{Tulane University, New Orleans, LA 70118, USA}
\affiliation{\GWU}
\affiliation{\NIST}
\author {R.A.~Arndt}
\affiliation{\GWU}
\author {W.J.~Briscoe}
\affiliation{\GWU}
\author {A.B.~Laptev} 
\altaffiliation[Present address: ]{Los Alamos National Laboratory, Los Alamos, NM 87545, USA}
\affiliation{\NIST}
\author {I.I.~Strakovsky}
\affiliation{\GWU}
\author {R.L.~Workman}
\affiliation{\GWU}

\pacs{11.80.Et,13.75.Cs,25.40.Cm,25.40.Dn}
\maketitle

\textbf{Abstract}
\textit{$-$ Absolute total cross sections for $np$ and
        $pp$ scattering below 1000~MeV are determined
        based on partial-wave analyses of $NN$
        scattering data.  These cross sections are
        compared with most recent ENDF/B and JENDL data
        files, and the Nijmegen partial-wave analysis.
        Systematic deviations from the ENDF/B and JENDL
        evaluations are found to exist in the low-energy
        region.}

\section{Introduction}
\label{sec:intro}

Nucleon-nucleon scattering is the simplest two-body 
reaction which allows an examination of different 
nuclear interaction models. Progress in the development 
of nuclear models is linked to the availability of 
high-quality data.  $np$ scattering is also used as a 
\textit{primary} standard in measurements 
neutron-induced nuclear reactions~\cite{stan,conf}. 
Its cross section is used in determining the flux of 
incoming neutrons.

This information is important in many applications, such 
as astrophysics [($n, \gamma$), ($n, \alpha$) and others] 
and the transmutation of nuclear waste [($n, f$), ($n, 
\gamma$) and others], energy generation and the
conceptual design of an innovative nuclear reactor being
carried out in the course of the Generation~IV 
initiative ~\cite{doe} [($n, f$), and neutron-actinoid 
elastic and inelastic scattering and others]. The 
increasing quality of neutron-induced nuclear reaction 
measurements requires a high-quality standard for $np$ 
cross sections, reproducing total $np$ cross sections 
with an accuracy of 1\% or better for energies below 
20~MeV~\cite{stan,sa04}.  The need for neutron data 
above 20~MeV up to hundreds of MeV with accuracy better 
than 10\%~\cite{sa04} leads to the requirement of cross 
section data for the $np$ \textit{reference} reaction 
with uncertainties at the few percent level.

An extensive database exists for nucleon-nucleon 
scattering, with measurements from laboratories 
worldwide.  These datasets, from the various laboratories, 
have different statistical and systematical uncertainties 
which must be taken into account when combined into a 
single fit.  At present, there are several evaluations 
of total $np$ cross sections below 20~MeV. Perhaps most 
widely known are the ENDF/B~\cite{ref} and JENDL
\cite{ref1} nuclear data files. An R-matrix analysis 
of the NN system~\cite{lanl} was used in course of the 
ENDF/B evaluation of $np$ cross sections whereas, in 
the JENDL $np$ total cross section evaluation, a 
method based on phase-shift data~\cite{sm86,sp82} was 
used.

Here we will concentrate on total $np$ and $pp$ cross 
section determined on the basis of recent energy-dependent 
(global) fits and associated single-energy solutions (SES) 
from the George Washington (GW) Data Analysis Center
\cite{sp07}.  Precise measurements collected over many 
years have helped to isolate descrepancies between 
different experiments and have contributed to a good 
description of nucleon-nucleon scattering at the level of 
both observables and amplitudes.

In Section~\ref{sec:expt}, we comment on the $np$ and $pp$ 
data which are available in the GW database and which have 
been used in this analysis.  In Section~\ref{sec:def}, we 
give a brief overview of the method used to fit data and 
extract amplitudes.  Then, in Section~\ref{sec:fit}, we 
present total $np$ and $pp$ cross sections determined on 
the basis of both global and SES results.  Finally, in 
Section~\ref{sec:conc}, we summarize our findings.

\section{Database}
\label{sec:expt}

The GW fit to $NN$ elastic scattering data covers an 
energy range from threshold up to a laboratory kinetic 
energy of 1300~MeV (for $np$ data) and 3000~MeV (for 
$pp$ data). The $np$ analysis was restricted to 1300~MeV 
due to a lack of high-energy data.  The full database 
includes all available unpolarized and polarized 
measurements.  A number of fits, from the GW group and 
others, are available through the on-line SAID facility
\cite{SAID}.

The evolution of the SAID database is summarized in 
Table~\ref{tab:tbl1}.  Over the course of seven previous 
GW $NN$ partial-wave analyses (PWA)
\cite{sp07,sp00,sm97,sm94,fa91,sm86,sp82}, the upper 
energy limit has been extended from 1200 (1100) to 3000 
(1300)~MeV in the laboratory proton (neutron) kinetic 
energy. (These fits are regularly updated online, and 
the designation SM97, for example, denotes the time of 
update - summer 1997.)

Not all of the available data have been used in each 
fit.  Some data with very large $\chi^2$ contributions 
have been excluded.  Redundant data are also excluded.  
These include total elastic cross sections based on 
differential cross sections already contained in the 
database.  Polarized measurements with uncertainties 
greater than 0.2 are not included as they have little 
influence in GW fits. However, all available data 
have been retained in the database (the excluded data 
labeled as ``flagged") so that comparisons can be made 
through the GW on-line facility~\cite{SAID}.  A complete 
description of the database, and those data not 
included in GW fits, is available from the authors
\cite{SAID}.

\section{Partial-Wave Analysis}
\label{sec:def}

Simultaneous fits to the full database are possible 
within the formalism used and described in previous 
GW analyses~\cite{sm86,sm94,sm97}. The observables
are represented in terms of partial-wave amplitudes,
using a Chew-Mandelstam K-matrix approach, which
incorporates the effect of an $N \Delta$ channel on 
the $NN$ scattering process. By parameterizing the 
K-matrix elements as functions of energy, data from 
threshold to 3000~MeV can be fitted simultaneously 
(both $pp$ and $np$, with a 1300~MeV limit for $np$).  
In general, GW partial-wave analyses have attempted 
to remain as model-independent as possible. The 
present (SP07) and previous energy-dependent 
solutions are compared in Table~\ref{tab:tbl1}.
 
In fitting the data, systematic uncertainty has 
been used as an overall normalization factor for 
angular distributions.  With each angular distribution, 
we associate the pair $(X,\epsilon_X)$: a normalization 
constant $(X)$ and its uncertainty $(\epsilon_X)$.  The 
quantity $\epsilon_X$ is generally associated with the
systematic uncertainty (if known).  The modified
$\chi^2$ function, to be minimized, is then given by
\begin{eqnarray}
\chi^2 = \sum_i\left(
{{X\theta_i - \theta^{\rm exp}_i}\over{\epsilon_i}}
\right)^2 + \left( {{X-1}\over{\epsilon_X}} \right)^2,
\nonumber\end{eqnarray}
where the subscript $i$ labels data points within the 
distribution, $\theta^{\rm exp}_i$ is an individual 
measurement, $\theta_i$ is the calculated value, and 
$\epsilon_i$ is the statistical uncertainty.  For 
total cross sections and excitation data, we have 
combined statistical and systematic uncertainties 
in quadrature.  Renormalization freedom significantly 
improves GW best-fit results, as shown in 
Table~\ref{tab:tbl2}. 

Starting from this global fit, we have also generated 
a series of SES results. Each SES is based on a 'bin' 
of scattering data spanning a narrow energy range. A 
total 43 SES have been generated, with central energy 
values ranging from 5~MeV to 2830~MeV, and bin widths 
varying from 2 to 75~MeV. For example, solution c5 is 
a fit to data between 4 and 6~MeV. In generating the 
SES, a linearized energy dependence is taken over the 
energy range, reducing the number of searched 
parameters. A systematic deviation between the SES 
and global fits is an indication of missing structure 
in the global fit (or possibly problems with a 
particular dataset). An error matrix is generated in 
the SES fits, which can be used to estimate the 
overall uncertainty in the global fit. The global 
and SES fit results, up to 1000~MeV, are summarized 
in Table~\ref{tab:tbl3}  (further details are given 
in Ref.~\cite{sp07}). 

\section{Total $np$ and $pp$ Cross Sections}
\label{sec:fit}

Isovector and isoscalar partial-wave amplitudes, 
determined through the partial-wave analysis, have 
been used to generate total $np$ and $pp$ cross 
sections presented in Tables~\ref{tab:tbl4} and 
\ref{tab:tbl5}. In Table~\ref{tab:tbl4}, we compare 
results from two global fits, SP07 (with an energy 
limit of 3000~MeV) and LE08 (a low-energy fit to 
25~MeV which searches 19 parameters, scattering 
length $a$ and effective range $r$ for 3 S-waves 
and 13 leading parameters for S, P, and D waves).  
LEO8 results in a $\chi^2$/data = 696/391 for $pp$ 
and 627/631 for $np$.  The SP07, LE08, and SES 
results below 20~MeV, are summarized in 
Table~\ref{tab:tbl6}.  Errors for LE08 have been 
generated from the error matrix and require some 
comments.
 
In the region below 25~MeV, there are numerous total 
cross section measurements for $np$, but not for 
$pp$ which is hindered by large Coulomb effects. As 
a result, the $np$ error estimates are more reasonable. 
Those quoted for $pp$ are far too small (lower limits) 
in the threshold region. In two cases (c5 and c150 of
Table~\ref{tab:tbl3}), the $\chi^2$ from the 
energy-dependent fit was actually lower than that 
from the SES fit.  There is a possibility that this 
can indicate that the energy bin for the SES fit is 
too broad.  But a rather narrow bin will not allow 
a stable result because of the lack of data to 
constrain the solution.

For the region above 25~MeV, in Table~\ref{tab:tbl5}, 
we compare the global fit SP07 with the grid of SES 
fits. Here the SES errors give a more accurate 
estimate of the uncertainty in our cross sections.  
The amplitudes found in GW fits to 1000~MeV have 
remained stable against the addition of new measurements 
for many years. Sufficient observables exist for a 
direct amplitude reconstruction at many energies, and 
we have compared GW amplitudes to those found in this 
way in Ref.~\cite{sp07}.

\begin{figure}[p]
\centerline{
\includegraphics[height=0.45\textwidth, angle=90]{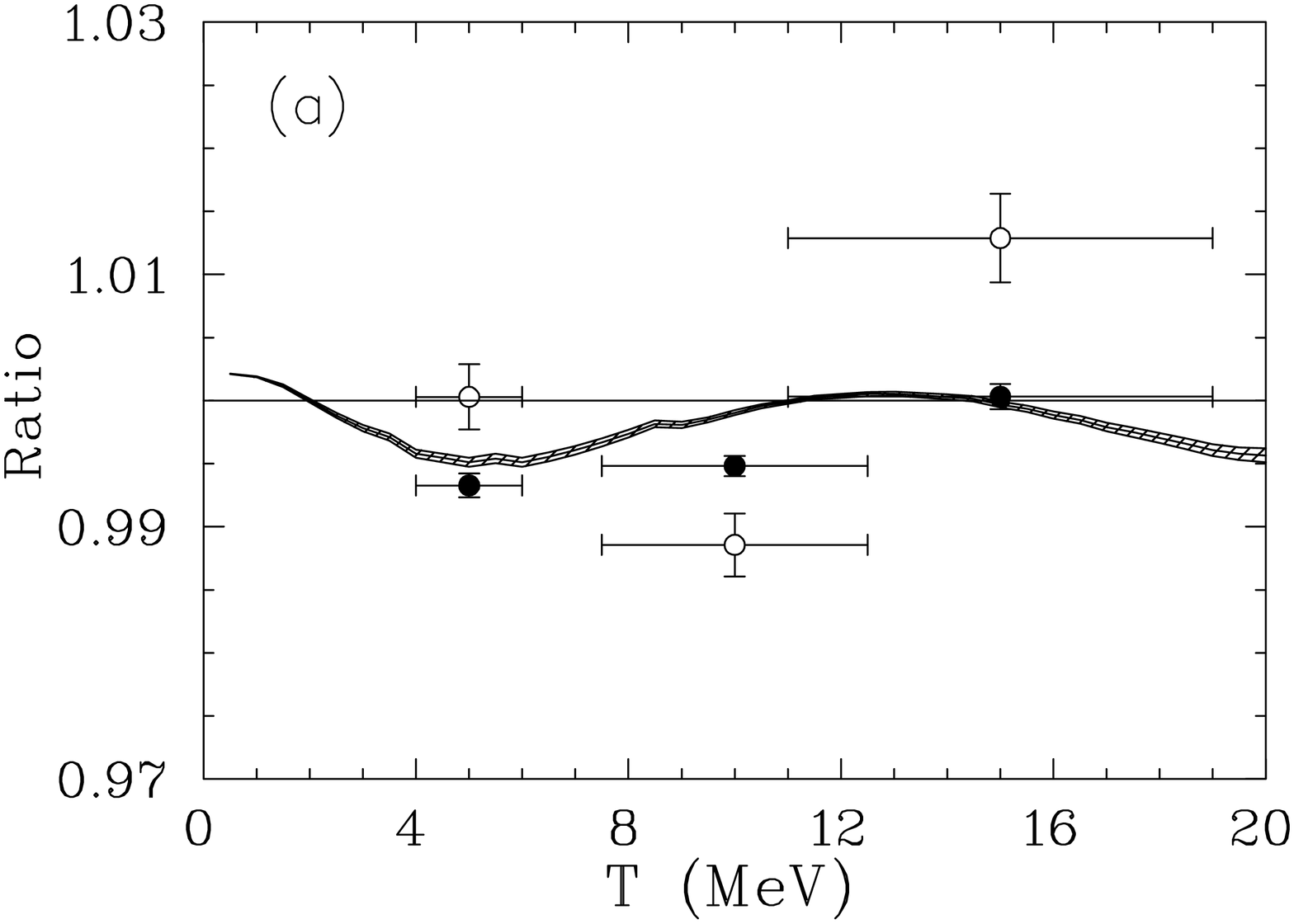}\hfill
\includegraphics[height=0.45\textwidth, angle=90]{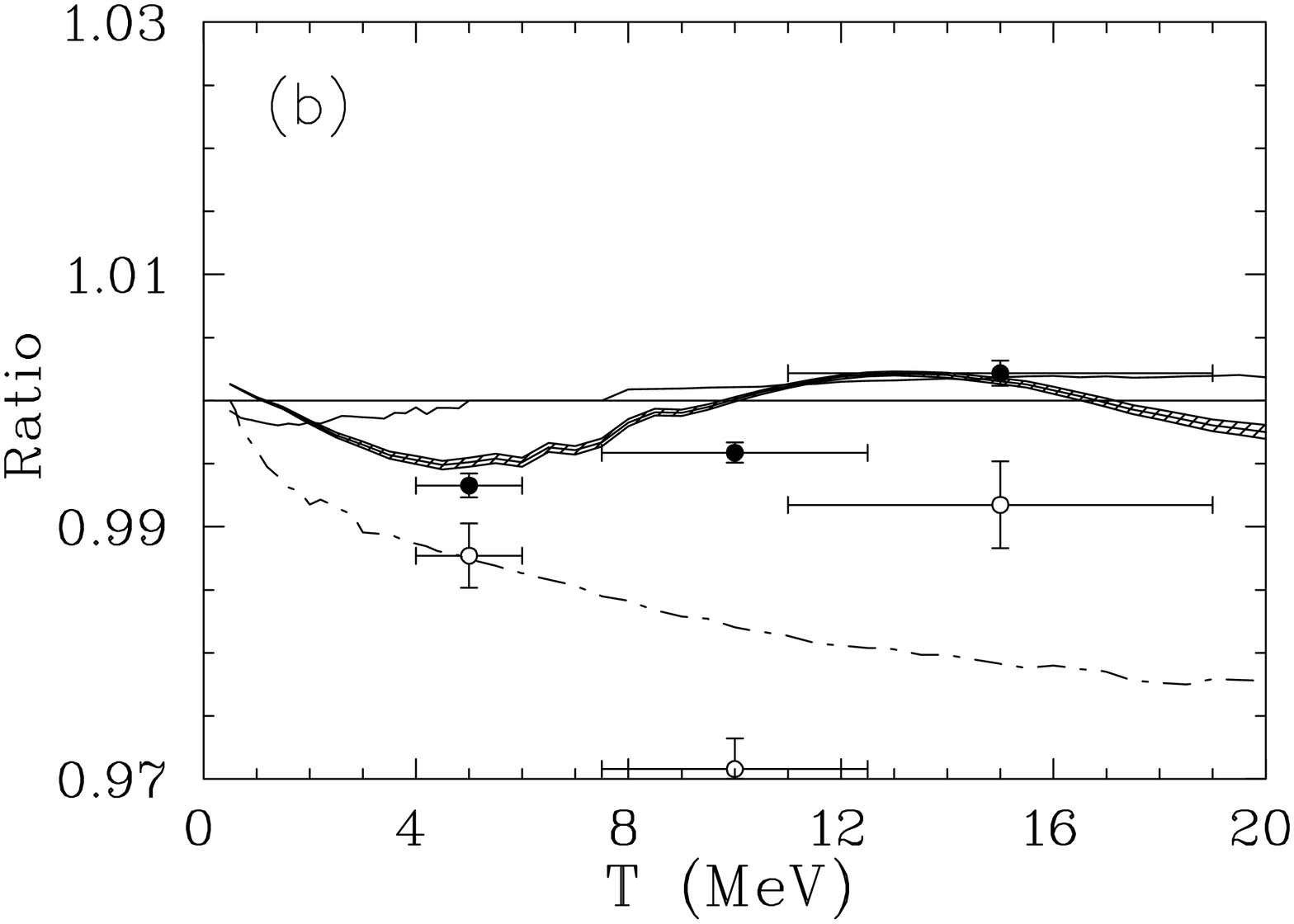}}
\vspace{3mm}
\caption{Ratios of total $np$ and $pp$ cross sections
         below 20~MeV. Horizontal bars give the
         energy binning of SES (Table~\ref{tab:tbl3}).
         (a) Single-energy to energy-dependent SP07
             \protect\cite{sp07} ratios are plotted.  $np$
             ($pp$) results are shown as solid (open)
             circles.  The band represents the ratio of 
             LE08 to SP07 for the $np$ case.
         (b) SES (solid circles for $np$ and open
             circles for $pp$) and SP07 (solid line for
             $np$ and dash-dotted line for $pp$) divided
             by Nijmegen PWA predictions
             \protect\cite{ni93} are plotted. The band             
             represents the ratio of LE08 to Nijmegen PWA    
             for the $np$ case. \label{fig:g1}}
\end{figure}

As cross sections change rapidly near threshold, we 
have chosen to display the agreement between various 
fits in terms of ratios. This gives a clear picture 
of the overall consistency and reveals cases where 
systematic deviations are present. The ratios of SES 
values to the global fit SP07, below 20~MeV, are 
displayed in Fig.~\ref{fig:g1}(a). Also plotted is 
a band showing the ratio of LE08 to SP07 
determinations of the $np$ cross section. As expected, 
this band more closely reproduces the $np$ SES, 
plotted as single points with error bars, than the 
3000~MeV fit SP07. Deviations are within 1\% for the 
$np$ determinations, and within 2\% for $pp$.    

In Fig.~\ref{fig:g1}(b), we plot ratios of SP07 and 
SES, for both $np$ and $pp$ cases, to the Nijmegen 
PWA predictions~\cite{ni93}.  The low-energy Nijmegen 
total $pp$ cross sections are systematically above 
SP07 (about 2\% or less) while $np$ cross sections 
agree with SP07 at better than the 0.3\% level.

\begin{figure}[p]
\centering{
\includegraphics[height=0.45\textwidth, angle=90]{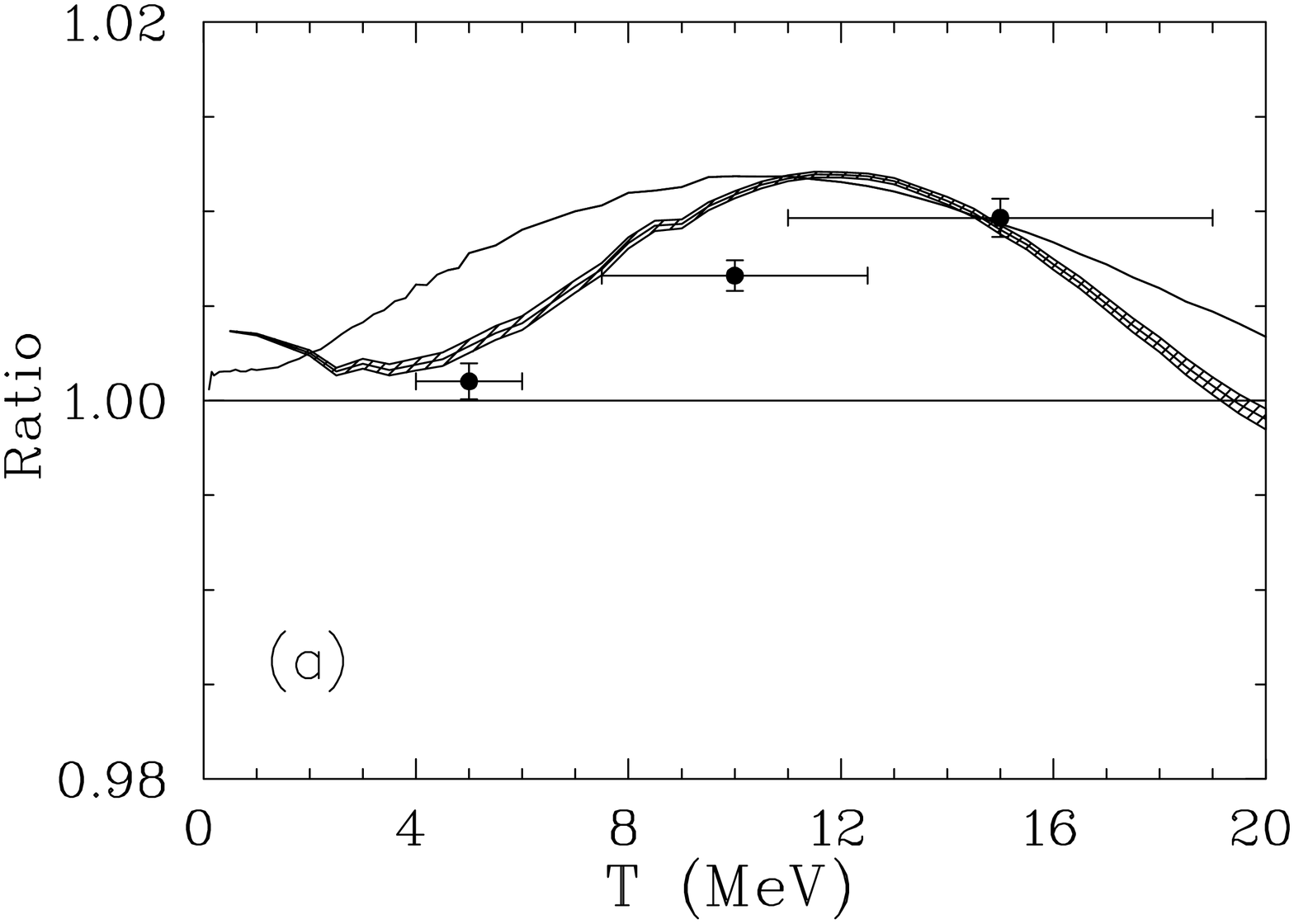}\hfill
\includegraphics[height=0.45\textwidth, angle=90]{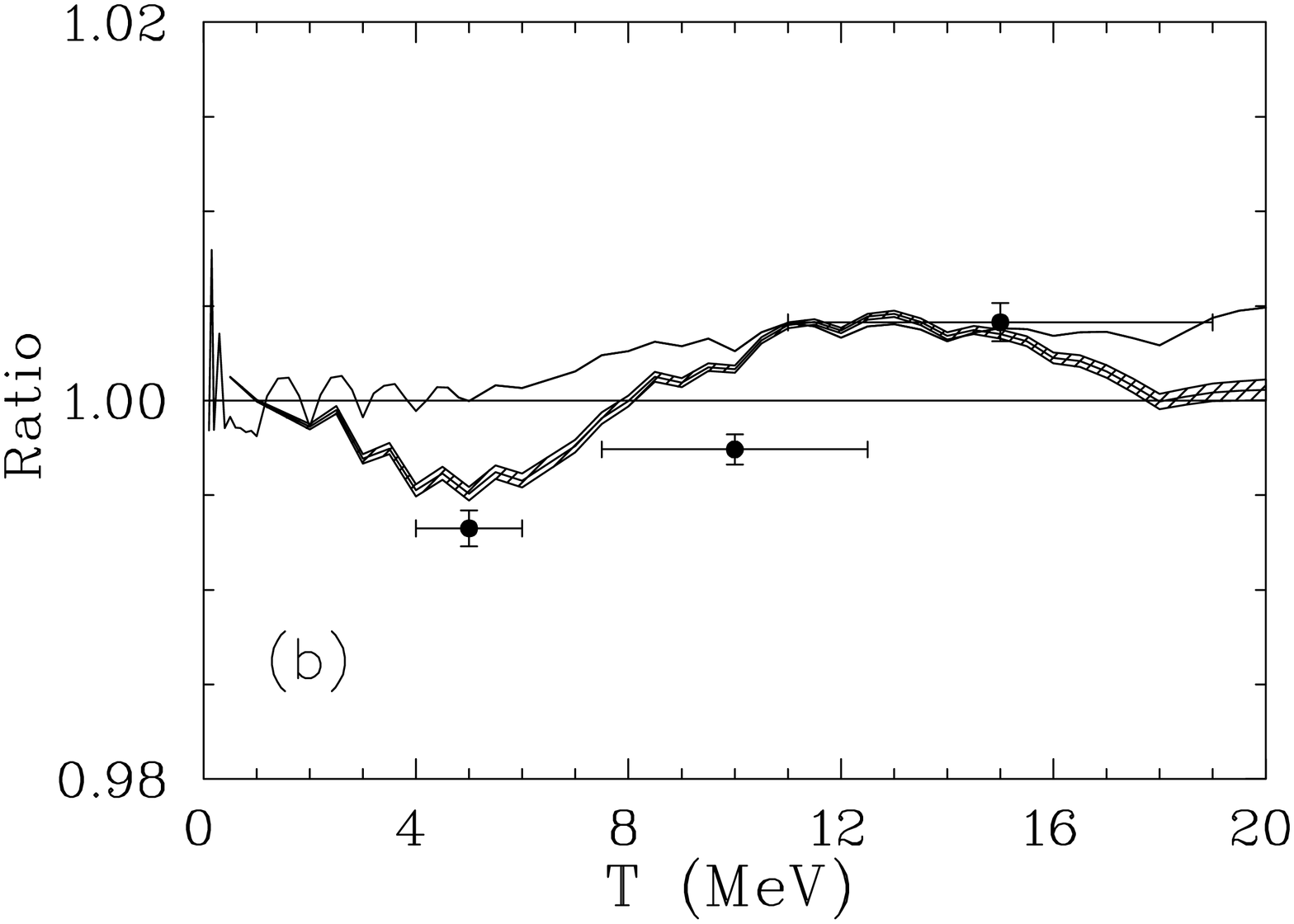}}
\vspace{3mm}
\caption{Ratio of total $np$ cross sections below 20~MeV.
         Horizontal bars give the energy binning of    
         SES (Table~\ref{tab:tbl3}).
         (a) Single-energy (solid circles) and SP07 (solid
             line) fits~\protect\cite{sp07} divided by the
             ENDF/B~\protect\cite{ref} results are plotted.
             The band gives a ratio of LE08 to ENDF/B.
         (b) The same for JENDL~\protect\cite{ref1}
             evaluated data. The wiggles seen in     
             Fig.~\protect\ref{fig:g2}(b) reflect a
             lack of smoothness in JENDL (SP07 and
             LE08 are a smooth function of energy.
             \label{fig:g2}}
\end{figure}

In Fig.~\ref{fig:g2}, we plot ratios of the GW $np$ 
fits with the ENDF/B and JENDL nuclear data files
\cite{ref,ref1}. A slightly better agreement is 
found with JENDL~\cite{ref1} than with ENDF/B
\cite{ref}, though the wiggles seen in 
Fig.~\ref{fig:g2}(b) reflect a lack of smoothness 
in JENDL (SP07 and LE08 are a smooth function of 
energy). The ENDF/B result is systematically below 
SP07 and the Nijmegen fit~\cite{ni93}, but the 
maximal deviation is only 1\%. SP07 and JENDL agree 
at the 0.5\% level over most of the region below 
20~MeV.

\begin{figure}[p]
\centerline{
\includegraphics[height=0.45\textwidth, angle=90]{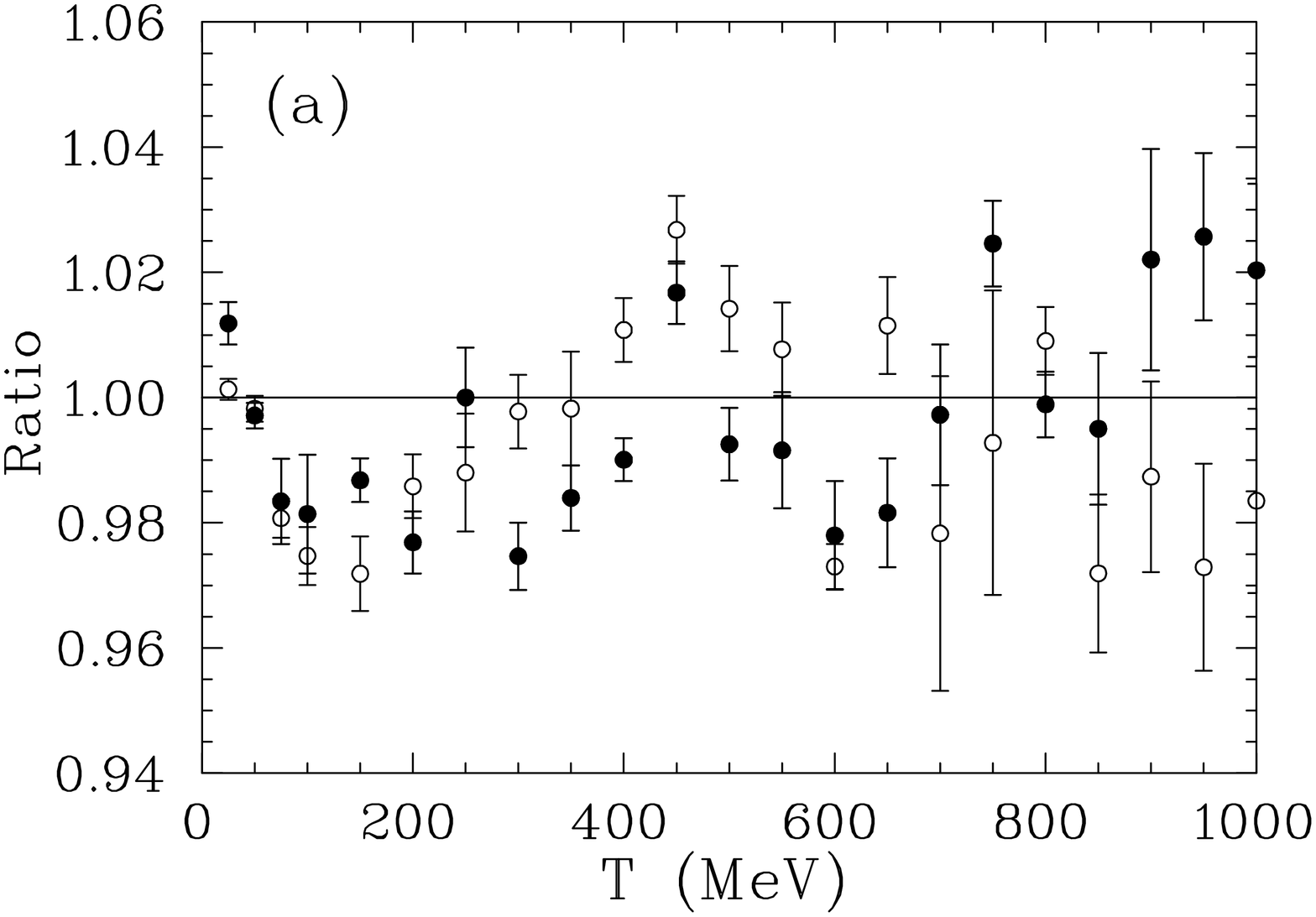}\hfill
\includegraphics[height=0.45\textwidth, angle=90]{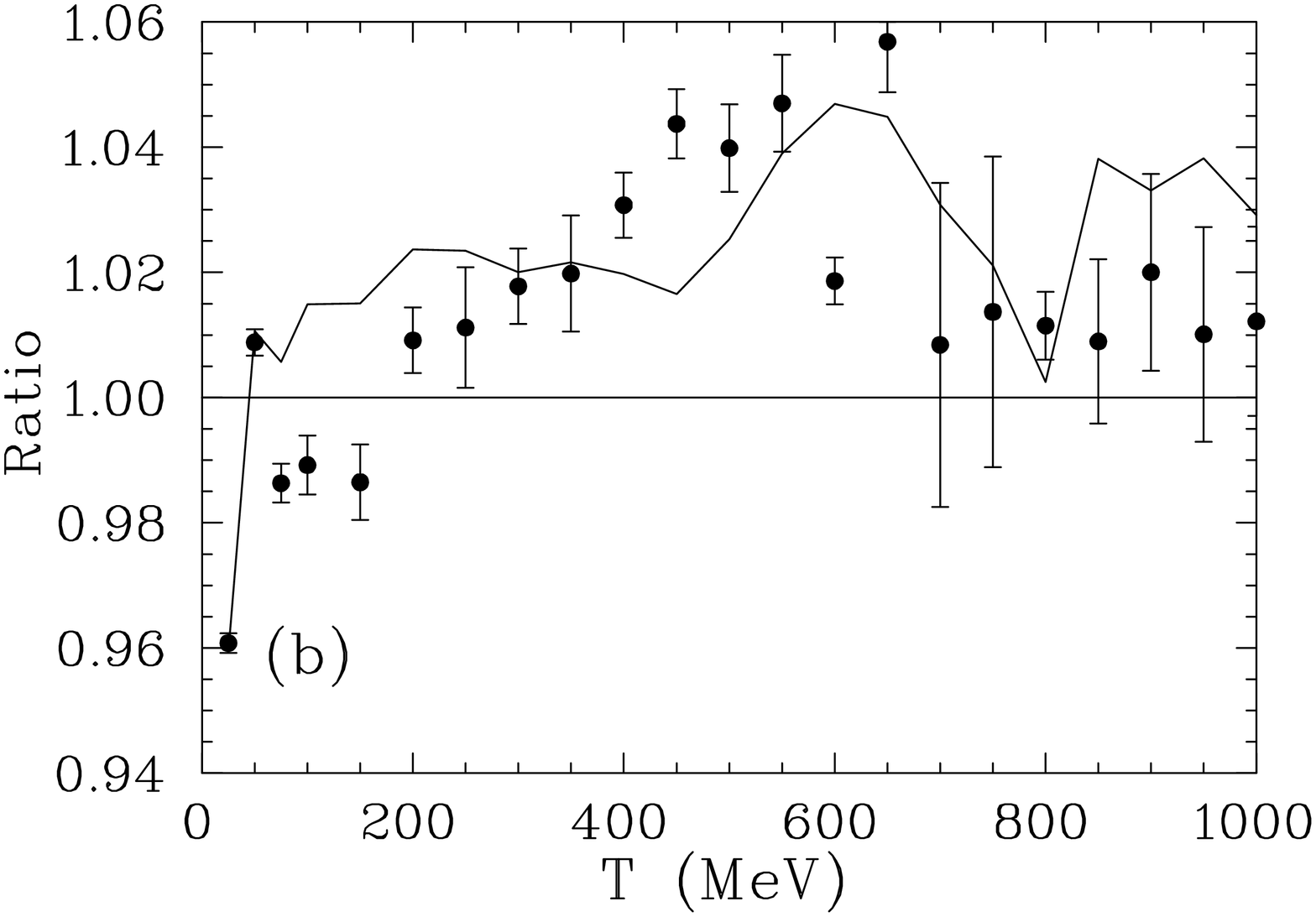}}
\vspace{3mm}
\caption{Ratios of total $np$ and $pp$ cross sections
         between 20 and 1000~MeV.
         (a) SES to SP07~\protect\cite{sp07} ratios
             are plotted. $np$ ($pp$) results are shown
             as solid (open) circles.
         (b) $np$ SES (solid circles) and SP07 (solid
             line) divided by JENDL/HE~\protect\cite{ref1}
             results are plotted. \label{fig:g3}}
\end{figure}

At higher energies (up to 1000~MeV), ratios of the 
grid of SES to SP07 differ from unity by less than 
3\% [Fig.~\ref{fig:g3}(a)]. Above 180~MeV, SAID $np$ 
cross sections are larger than JENDL/HE~\cite{ref1} 
by up to 5\% [Fig.~\ref{fig:g3}(b)].

\section{Summary and Conclusions}
\label{sec:conc}

We have generated fits to describe the total $np$ and 
$pp$ scattering cross sections below 1000~MeV. These 
fits have been both energy dependent (SP07, LE08) and 
single-energy (analyzing narrow bins of data). The 
uncertainties associated  with our total $np$ cross 
sections below 20~MeV are clearly less than 1\%. The 
agreement between SP07, JENDL, and the Nijmegen 
analysis, suggests an uncertainty of 0.5\% or less.
A comparison with ENDF/B shows deviations of 1\% or
less.  Errors on the LE08 solution in 
Table~\ref{tab:tbl4}, while obtained using a 
well-defined method, are lower bounds as they do not 
account for systematics effects.

For the $pp$ cross sections, uncertainties are 
larger and systematic disagreements are evident in 
comparisons with the Nijmegen PWA. The main problem 
stems from a lack of relevant $pp$ data at low 
energies. Here also, at low energies, the various 
determinations agree at the few-percent level. 

The advantage of the GW parameterization is its 
smooth energy dependence and coverage from threshold 
to high energies. We also have the capability to 
modify the GW fits to either generate SES centered 
on a particular energy, or produce lower-energy fits 
when a specific energy region is of interest. We 
will continue to update both GW energy-dependent 
and single-energy solutions as the new measurements 
become available.

\vspace{5mm}
\acknowledgments

The authors express their gratitude to A.D.~Carlson, 
G.M.~Hale, and T.~Hill for the useful discussion.  
This work was supported in part by the U.S.~Department 
of Energy under Grant DE--FG02--99ER41110.  The 
authors acknowledge partial support from Jefferson 
Lab, by the Southeastern Universities Research 
Association under DOE contract DE--AC05--84ER40150.

\newpage


\newpage
\begin{table}[th]
\caption{Comparison of recent SP07~\protect\cite{sp07}
         and previous 
         SP00~\protect\cite{sp00},
         SM97~\protect\cite{sm97},
         SM94~\protect\cite{sm94},
         FA91~\protect\cite{fa91},
         SM86~\protect\cite{sm86}, and
         SP82~\protect\cite{sp82} GW energy-dependent
         partial-wave analyses.  The $\chi^2$ values 
         for previous solutions correspond to
         published results. \label{tab:tbl1}}
\vspace{2mm}
\begin{tabular}{|c|c|c|c|c|}
\colrule
Solution & Range & $\chi^2$/$np$ data & Range & $\chi^2$/$pp$ data \\
         & (MeV) &                    & (MeV) &                    \\
\colrule
SP07     &0--1300&    21496/12693     &0--3000&    44463/24916     \\
SP00     &0--1300&    18693/11472     &0--3000&    36617/21796     \\
SM97     &0--1300&    17437/10854     &0--2500&    28686/16994     \\
SM94     &0--1300&    17516/10918     &0--1600&    22371/12838     \\
FA91     &0--1100&    13711/ 7572     &0--1600&    20600/11880     \\
SM86     &0--1100&     8871/ 5474     &0--1200&    11900/ 7223     \\
SP82     &0--1100&     9103/ 5283     &0--1200&     9199/ 5207     \\
\colrule
\end{tabular}
\end{table}
\newpage
\begin{table}[th]
\caption{$\chi^2$/data ($pp$ and $np$) normalized
        (Norm) and unnormalized (Unnorm) below
        1000~MeV for GW fits SP07~\protect\cite{sp07},
        SP00~\protect\cite{sp00},
        SM97~\protect\cite{sm97}, and the
        Nijmegen PWA fit~\protect\cite{ni93}
        (the Nijmegen solution is valid up to 350~MeV
        - the final tabulated energy range is
        200--350~MeV). Last two columns list numbers
        of $np$ and $pp$ data included in fit SP07
        and its associated SES. \label{tab:tbl2}}
\vspace{2mm}
\begin{tabular}{|c|c|c|c|c|c|c|}
\colrule
 Range   &    SP07     &   SP00     &    SM97    & Nijmegen93 & Data & Data \\
 (MeV)   & Norm/Unnorm & Norm/Unnorm& Norm/Unnorm& Norm/Unnorm& $np$ & $pp$ \\
\colrule
  0--   4&  2.5 / 28.0 & 2.5 / 28.0 & 2.5 / 27.9 & 3.3 / 26.7 &   63 &  193 \\
  0--  20&  1.8 / 13.2 & 1.9 / 13.3 & 2.3 / 13.8 & 2.9 / 10.1 &  468 &  389 \\
  0-- 200&  1.5 /  7.2 & 1.5 /  7.1 & 1.7 /  6.8 & 1.7 /  5.9 & 2381 & 1491 \\
200-- 400&  1.3 /  3.3 & 1.3 /  3.3 & 1.4 /  3.3 & 1.3 /  2.5 & 2208 & 2172 \\
400-- 600&  1.5 /  8.7 & 1.4 /  8.0 & 1.5 / 10.6 &            & 2779 & 3635 \\
600-- 800&  1.5 /  8.4 & 1.5 /  7.7 & 1.4 / 10.6 &            & 2529 & 3974 \\
800--1000&  1.4 /  3.4 & 1.4 /  3.4 & 1.4 /  3.1 &            & 2112 & 3274 \\
\colrule
\end{tabular}
\end{table}
\newpage
\begin{table}[th]
\caption{Single-energy (binned) fits~\protect\cite{sp07}
         of combined elastic $np$ and $pp$ scattering
         data below 1000~MeV. $\chi^2_E$ is given by the
         energy-dependent fit, SP07~\protect\cite{sp07},
         over the same energy intervals. \label{tab:tbl3}}
\vspace{2mm}
\begin{tabular}{|c|c|c|c|c|c|}
\colrule
Solution & Range  &   $pp$    &              &    $np$    &               \\
         &  (MeV) &$\chi^2_E$ & $\chi^2$/data& $\chi^2_E$ & $\chi^2$/data \\
\colrule
c5   &   4$-$   6 &  37 &  22/  28 &  74 &  77/ 62 \\
c10  &   7$-$  12 & 128 &  81/  88 & 255 & 222/106 \\
c15  &  11$-$  19 &  55 &  16/  27 & 366 & 219/247 \\
c25  &  19$-$  30 & 251 & 120/ 114 & 319 & 293/316 \\
c50  &  32$-$  67 & 331 & 283/ 224 & 721 & 660/467 \\
c75  &  60$-$  90 &  61 &  48/  72 & 652 & 516/355 \\
c100 &  80$-$ 120 & 155 & 149/ 154 & 521 & 437/389 \\
c150 & 125$-$ 174 & 352 & 311/ 287 & 533 & 542/352 \\
c200 & 175$-$ 225 & 570 & 528/ 435 & 870 & 742/519 \\
c250 & 225$-$ 270 & 289 & 266/ 263 & 601 & 543/438 \\
c300 & 276$-$ 325 & 853 & 815/ 805 &1163 &1074/770 \\
c350 & 325$-$ 375 & 522 & 503/ 474 & 509 & 462/381 \\
c400 & 375$-$ 425 & 819 & 771/ 680 &1295 &1208/805 \\
c450 & 425$-$ 475 &1307 &1186/ 877 &1216 &1194/912 \\
c500 & 475$-$ 525 &1811 &1556/1215 &1386 &1240/815 \\
c550 & 525$-$ 575 &1075 & 938/ 817 &1192 & 987/719 \\
c600 & 575$-$ 625 &1196 &1045/ 838 & 517 & 423/367 \\
c650 & 625$-$ 675 & 980 & 959/ 807 &1502 &1265/877 \\
c700 & 675$-$ 725 & 982 & 956/ 887 & 460 & 396/386 \\
c750 & 725$-$ 775 &1474 &1085/ 926 & 545 & 508/382 \\
c800 & 775$-$ 824 &2056 &1833/1385 &1777 &1427/950 \\
c850 & 827$-$ 874 &1296 &1142/ 980 & 500 & 393/365 \\
c900 & 876$-$ 924 & 542 & 406/ 444 & 936 & 730/627 \\
c950 & 926$-$ 974 & 961 & 769/ 728 & 528 & 356/353 \\
c999 & 976$-$1020 &1206 &1010/ 776 & 421 & 274/329 \\
\colrule
\end{tabular}
\end{table}
\newpage
\begin{table}[th]
\caption{Comparison of SP07~\protect\cite{sp07} 
         and LE08 results for total $np$ ($\sigma^n$)
         and $pp$ ($\sigma^p$) cross sections below
         25~MeV. See text for a discussion of
         uncertainties. \label{tab:tbl4}}
\vspace{2mm}
\begin{tabular}{|c|c|c|c|c|c|c|c|c|c|c|}
\colrule
  T  &      LE08     &      SP07     &      LE08     &      SP07     &&  T  &      LE08
     &       SP07    &      LE08     &      SP07     \\
(MeV)&$\sigma^p$~(mb)&$\sigma^p$~(mb)&$\sigma^n$~(mb)&$\sigma^n$~(mb)&&(MeV)&$\sigma^p$~(mb)&$\sigma^p$~(mb)&$\sigma^n$~(mb)&$\sigma^n$~(mb)\\
\colrule
 0.5 & 1100.00$\pm$0.01 & 1098.00 & 6148.00$\pm$0.07 & 6135.00 &&
13.0 &  264.00$\pm$0.10 &  263.70 &  745.40$\pm$0.12 &  745.00\\
 1.0 & 1513.00$\pm$0.01 & 1513.00 & 4261.00$\pm$0.19 & 4253.00 &&
13.5 &  253.60$\pm$0.11 &  253.00 &  719.10$\pm$0.13 &  718.80\\
 1.5 & 1469.00$\pm$0.01 & 1471.00 & 3421.00$\pm$0.32 & 3417.00 &&
14.0 &  243.90$\pm$0.13 &  243.10 &  694.40$\pm$0.14 &  694.20\\
 2.0 & 1323.00$\pm$0.01 & 1325.00 & 2911.00$\pm$0.44 & 2911.00 &&
14.5 &  234.90$\pm$0.14 &  233.80 &  671.10$\pm$0.14 &  671.00\\
 2.5 & 1172.00$\pm$0.01 & 1175.00 & 2555.00$\pm$0.52 & 2558.00 &&
15.0 &  226.60$\pm$0.15 &  225.20 &  649.00$\pm$0.15 &  649.20\\
 3.0 & 1040.00$\pm$0.01 & 1042.00 & 2286.00$\pm$0.57 & 2291.00 &&
15.5 &  218.80$\pm$0.16 &  217.10 &  628.20$\pm$0.16 &  628.60\\
 3.5 &  927.70$\pm$0.01 &  930.60 & 2073.00$\pm$0.61 & 2079.00 &&
16.0 &  211.60$\pm$0.17 &  209.60 &  608.40$\pm$0.17 &  609.10\\
 4.0 &  833.90$\pm$0.01 &  836.70 & 1898.00$\pm$0.61 & 1906.00 &&
16.5 &  204.80$\pm$0.18 &  202.50 &  589.70$\pm$0.18 &  590.60\\
 4.5 &  755.00$\pm$0.01 &  757.70 & 1752.00$\pm$0.61 & 1760.00 &&
17.0 &  198.50$\pm$0.20 &  195.80 &  571.90$\pm$0.19 &  573.10\\
 5.0 &  688.10$\pm$0.01 &  690.60 & 1627.00$\pm$0.59 & 1635.00 &&
17.5 &  192.60$\pm$0.21 &  189.50 &  555.00$\pm$0.20 &  556.40\\
 5.5 &  630.90$\pm$0.01 &  633.30 & 1519.00$\pm$0.55 & 1526.00 &&
18.0 &  187.00$\pm$0.23 &  183.60 &  539.00$\pm$0.21 &  540.60\\
 6.0 &  581.60$\pm$0.02 &  583.80 & 1424.00$\pm$0.52 & 1431.00 &&
18.5 &  181.80$\pm$0.24 &  178.00 &  523.70$\pm$0.23 &  525.50\\
 6.5 &  538.70$\pm$0.02 &  540.80 & 1341.00$\pm$0.47 & 1347.00 &&
19.0 &  176.90$\pm$0.25 &  172.80 &  509.20$\pm$0.24 &  511.20\\
 7.0 &  501.10$\pm$0.02 &  503.10 & 1266.00$\pm$0.43 & 1271.00 &&
19.5 &  172.30$\pm$0.27 &  167.80 &  495.40$\pm$0.25 &  497.50\\
 7.5 &  468.00$\pm$0.03 &  469.90 & 1199.00$\pm$0.37 & 1203.00 &&
20.0 &  167.90$\pm$0.28 &  163.00 &  482.30$\pm$0.26 &  484.40\\
 8.0 &  438.70$\pm$0.03 &  440.40 & 1139.00$\pm$0.33 & 1142.00 &&
20.5 &  164.00$\pm$0.30 &  158.70 &  469.80$\pm$0.28 &  471.90\\
 8.5 &  412.50$\pm$0.04 &  414.00 & 1084.00$\pm$0.29 & 1086.00 &&
21.0 &  160.20$\pm$0.32 &  154.40 &  457.90$\pm$0.29 &  460.00\\
 9.0 &  389.00$\pm$0.04 &  390.30 & 1033.00$\pm$0.25 & 1035.00 &&
21.5 &  156.50$\pm$0.33 &  150.40 &  446.50$\pm$0.31 &  448.50\\
 9.5 &  367.80$\pm$0.05 &  369.00 &  987.10$\pm$0.21 &  988.60 &&
22.0 &  153.10$\pm$0.35 &  146.50 &  435.70$\pm$0.32 &  437.60\\
10.0 &  348.60$\pm$0.06 &  349.60 &  944.60$\pm$0.18 &  945.50 &&
22.5 &  149.90$\pm$0.36 &  142.80 &  425.30$\pm$0.34 &  427.00\\
10.5 &  331.20$\pm$0.06 &  332.00 &  905.30$\pm$0.15 &  905.70 &&
23.0 &  146.80$\pm$0.38 &  139.30 &  415.50$\pm$0.36 &  417.00\\
11.0 &  315.40$\pm$0.07 &  316.00 &  868.80$\pm$0.13 &  868.90 &&
23.5 &  143.90$\pm$0.40 &  135.90 &  406.10$\pm$0.37 &  407.30\\
11.5 &  300.90$\pm$0.08 &  301.20 &  834.80$\pm$0.12 &  834.60 &&
24.0 &  141.20$\pm$0.41 &  132.70 &  397.10$\pm$0.39 &  398.00\\
12.0 &  287.50$\pm$0.09 &  287.70 &  803.00$\pm$0.12 &  802.70 &&
24.5 &  138.60$\pm$0.43 &  129.60 &  388.50$\pm$0.41 &  389.00\\
12.5 &  275.30$\pm$0.10 &  275.20 &  773.30$\pm$0.12 &  772.90 &&
25.0 &  136.20$\pm$0.45 &  126.70 &  380.30$\pm$0.43 &  380.40\\
\colrule
\end{tabular}
\end{table}
\newpage
\begin{table}[th]
\caption{Comparison of SES and SP07~\protect\cite{sp07}
         results for total $np$ ($\sigma^n$) and $pp$
         ($\sigma^p$) cross sections between 25 and
         1000~MeV. \label{tab:tbl5}}
\vspace{2mm}
\begin{tabular}{|c|c|c|c|c|c|c|}
\colrule
   T   &Solution&       SES     &       SP07    &       SES     &        SP07    
\\
 (MeV) &        &$\sigma^p$~(mb)&$\sigma^p$~(mb)&$\sigma^n$~(mb)& 
$\sigma^n$~(mb)\\
\colrule
  25.0 &   c25  & 128.20$\pm$0.43 & 126.70 & 380.90$\pm$0.63 & 380.40 \\
  50.0 &   c50  &  58.78$\pm$0.12 &  58.95 & 168.10$\pm$0.35 & 168.40 \\
  75.0 &   c75  &  39.14$\pm$0.27 &  39.80 & 101.70$\pm$0.32 & 103.70 \\
 100.0 &  c100  &  31.11$\pm$0.30 &  31.70 &  73.62$\pm$0.35 &  75.53 \\
 150.0 &  c150  &  25.40$\pm$0.09 &  25.74 &  50.77$\pm$0.31 &  52.24 \\
 200.0 &  c200  &  23.66$\pm$0.12 &  24.22 &  42.43$\pm$0.22 &  43.04 \\
 250.0 &  c250  &  23.89$\pm$0.19 &  23.89 &  37.89$\pm$0.36 &  38.35 \\
 300.0 &  c300  &  23.46$\pm$0.13 &  24.07 &  35.53$\pm$0.21 &  35.61 \\
 350.0 &  c350  &  24.53$\pm$0.13 &  24.93 &  34.05$\pm$0.31 &  34.11 \\
 400.0 &  c400  &  25.91$\pm$0.09 &  26.17 &  33.70$\pm$0.17 &  33.34 \\
 450.0 &  c450  &  28.53$\pm$0.14 &  28.06 &  34.13$\pm$0.18 &  33.24 \\
 500.0 &  c500  &  30.60$\pm$0.18 &  30.83 &  34.30$\pm$0.23 &  33.82 \\
 550.0 &  c550  &  34.13$\pm$0.32 &  34.42 &  35.24$\pm$0.26 &  34.97 \\
 600.0 &  c600  &  37.29$\pm$0.33 &  38.13 &  35.32$\pm$0.13 &  36.30 \\
 650.0 &  c650  &  40.54$\pm$0.36 &  41.30 &  37.86$\pm$0.29 &  37.43 \\
 700.0 &  c700  &  43.53$\pm$0.49 &  43.65 &  37.40$\pm$0.96 &  38.23 \\
 750.0 &  c750  &  46.23$\pm$0.31 &  45.12 &  38.38$\pm$0.94 &  38.66 \\
 800.0 &  c800  &  45.83$\pm$0.24 &  45.88 &  39.15$\pm$0.21 &  38.80 \\
 850.0 &  c850  &  45.97$\pm$0.56 &  46.20 &  37.71$\pm$0.49 &  38.80 \\
 900.0 &  c900  &  47.32$\pm$0.82 &  46.30 &  38.27$\pm$0.59 &  38.76 \\
 950.0 &  c950  &  47.49$\pm$0.62 &  46.30 &  37.69$\pm$0.64 &  38.74 \\
1000.0 &  c999  &  47.19$\pm$0.64 &  46.25 &  38.14$\pm$0.57 &  38.78 \\
\colrule
\end{tabular}
\end{table}
\newpage
\begin{table}[th]
\caption{Single-energy (binned) fits~\protect\cite{sp07}
         of combined elastic $np$ and $pp$ scattering
         data below 20~MeV. $\chi^2_E$ are given by
         the energy-dependent SP07~\protect\cite{sp07},
         LE08, and NI93~\protect\cite{ni93} fits, over
         the same energy intervals. Number of searched
         parameters in SES is given in the 2nd column
         in brackets. \label{tab:tbl6}}
\vspace{2mm}
\begin{tabular}{|c|c|c|c|c|c|}
\colrule
Solution & Range & SP07      & LE08     &   SES   & NI93          \\
         & (MeV) &$\chi^2_E$ &$\chi^2_E$&$\chi^2$ &$\chi^2_E$/data\\
\colrule
for $pp$ &          &     &     &     &         \\
c5   &  4 $-$  6~(6) &  37 &  30 &  22 &  51/28  \\
c10  &  7 $-$ 12~(6) & 129 &  98 &  81 & 107/88  \\
c15  & 11 $-$ 19~(8) &  55 &  37 &  17 &  15/27  \\
\colrule
for $np$ &          &     &     &     &         \\
c5   &  4 $-$  6~(6) &  75 &  58 &  78 &  78/ 62 \\
c10  &  7 $-$ 12~(6) & 256 & 138 & 222 & 132/106 \\
c15  & 11 $-$ 19~(8) & 366 & 231 & 219 & 246/247 \\
\colrule
\end{tabular}
\end{table}
\newpage
\newpage
\newpage
\end{document}